\documentclass[twocolumn]{pasj00}
\usepackage[dvips]{color}
\usepackage{multicol}

\begin{document}
\SetRunningHead{N. Narita et al.}{Radial Velocities and the Orbit of XO-2b}
\Received{2011/08/22}
\Accepted{2011/10/12}

\title{XO-2b: a Prograde Planet with a Negligible Eccentricity,\\
and an Additional Radial Velocity Variation$^*$}

\author{
Norio \textsc{Narita},\altaffilmark{1}
Teruyuki \textsc{Hirano},\altaffilmark{2}
Bun'ei \textsc{Sato},\altaffilmark{3}
Hiroki \textsc{Harakawa},\altaffilmark{3}\\
Akihiko \textsc{Fukui},\altaffilmark{4,5}
Wako \textsc{Aoki},\altaffilmark{1}
and Motohide \textsc{Tamura}\altaffilmark{1}
}

\altaffiltext{1}{
National Astronomical Observatory of Japan, 2-21-1 Osawa,
Mitaka, Tokyo, 181-8588, Japan
}

\altaffiltext{2}{
Department of Physics, The University of Tokyo, Tokyo, 113-0033, Japan
}

\altaffiltext{3}{
Department of Earth and Planetary Sciences, Tokyo Institute of Technology,
Tokyo, 152-8550, Japan
}

\altaffiltext{4}{
Okayama Astrophysical Observatory, National Astronomical Observatory,\\
3037-5 Honjo, Kamogata, Asakuchi, Okayama 719-0232, Japan
}

\altaffiltext{5}{
Solar-Terrestrial Environment Laboratory, Nagoya University,
Nagoya, 464-8601, Japan
}
\email{norio.narita@nao.ac.jp}


\maketitle

\begin{abstract}
We present precise radial velocities of XO-2 taken with the Subaru HDS,
covering two transits of XO-2b with an interval of nearly two years.
The data suggest that the orbital eccentricity of XO-2b is consistent with
zero within 2$\sigma$ ($e=0.045\pm0.024$) and the orbit of XO-2b is prograde
(the sky-projected spin-orbit alignment angle
$\lambda=10^{\circ}\pm72^{\circ}$).
The poor constraint of $\lambda$ is due to a small impact parameter
(the orbital inclination of XO-2b is almost 90$^{\circ}$).
The data also provide an improved estimate of the mass of XO-2b as
$0.62\pm0.02$~$M_{\rm Jup}$.
We also find a long-term radial velocity variation in this system.
Further radial velocity measurements are necessary to
specify the cause of this additional variation.
\end{abstract}
\footnotetext[*]{Based on data collected at Subaru Telescope,
which is operated by the National Astronomical Observatory of Japan.}

\section{Introduction}

Planetary orbits in binary systems provide useful clues
to learn planetary migration mechanisms in binary systems.
Planets in binary systems have a chance to take a different
path of orbital migration than planets in single star systems.
That is, due to the presence of a binary companion,
an inner planet can evolve into a highly eccentric and highly tilted,
or even retrograde orbit, by the mechanism known as the Kozai migration
\citep{1962AJ.....67..591K, 2003ApJ...589..605W,
2007ApJ...669.1298F,2011arXiv1106.3329L,2011arXiv1106.3340K}.
Observable evidences of the Kozai migration are large eccentricity
and large spin-orbit misalignment of an inner planet.
One can measure the orbital eccentricity via
radial velocity (RV) measurements, and one can also learn
the spin-orbit alignment angle at least in sky-projection
via the Rossiter-McLaughlin effect
(hereafter the RM effect: \cite{1924ApJ....60...15R},
\cite{1924ApJ....60...22M,
2005ApJ...622.1118O, 2010ApJ...709..458H};
Hirano~et~al.~2011a) in transiting systems.

One interesting example of a transiting planet in a wide
binary system is HD~80606b, which has a large eccentricity
($e=0.93$, \cite{2001A&A...375L..27N})
and a large sky-projected spin-orbit alignment angle
($\lambda=42^{\circ}$, \cite{2009A&A...498L...5M, 2009A&A...502..695P,
2009ApJ...703.2091W, 2010A&A...516A..95H}).
The orbit of HD~80606b has been well explained by
a scenario of the Kozai migration due to the presence of
the binary companion HD~80607
\citep{2003ApJ...589..605W}.
Another interesting example is WASP-8b, which has an eccentric and
retrograde orbit with a wide binary companion
\citep{2010A&A...517L...1Q},
although a Kozai migration scenario for WASP-8b has not
been well examined.
Other transiting planets in binary systems, however,
show neither large eccentricity nor 
large spin-orbit misalignment
(e.g., HD189733b: \cite{2006ApJ...641L..57B, 2006ApJ...653L..69W},
TrES-4b: \cite{2007ApJ...667L.195M, 2010PASJ...62..653N}).
This fact can be understood by considering that the Kozai migration
needs to meet some stringent conditions (see e.g.,
\cite{1997AJ....113.1915I, 2007ApJ...670..820W, 2010PASJ...62..779N}).
To learn the occurrence frequency of the Kozai migration
for planets in binary systems, it is important to
increase the number of observations for such transiting systems.

We here report on the orbit of the transiting planet XO-2b,
which was discovered by \citet{2007ApJ...671.2115B}
in the course of the XO survey.
The host star XO-2 is a relatively faint ($V=11.18$) K0V star
(the stellar mass $0.98\pm0.02 M_{\rm s}$ and the stellar radius
$0.97\pm0.02 R_{\rm s}$) at $\sim$150 pc from the Sun
\citep{2007ApJ...671.2115B}.
XO-2 has the twin (the same spectral type) stellar companion
XO-2S at 31'' ($\sim$4600AU) separation.
We note that the Kozai migration for XO-2b could occur if the eccentricity
of the binary orbit (currently unknown) is very large ($e\ge0.9$),
when the timescale of the Kozai migration is shorter than the timescale
of a perturbation due to General Relativity for XO-2b
\citep{2007ApJ...670..820W}.
\citet{2007ApJ...671.2115B} reported the mass, radius, orbital period
of XO-2b as $0.57\pm0.06~M_{\rm Jup}$, $0.98^{+0.03}_{-0.01}~R_{\rm Jup}$,
and $P=2.615857 \pm 0.000005$ days, respectively.
The eccentricity of XO-2b was assumed to be zero in the discovery paper,
and no measurement of the RM effect in this system has been reported.
In this letter, we present RVs of XO-2 spanning about two years and
covering two full transits of XO-2b allowing us to model the RM effect.
Our RV data suggest a small eccentricity and
a likely spin-orbit alignment of XO-2b,
concluding no supporting evidence of the Kozai migration.
In addition, we find a long-term RV trend for XO-2.
This RV acceleration may suggest a third body in this system.

The rest of this letter is organized as follows.
We present details of our observations and reductions of
RV data in section~2.
We describe our model to fit the observed data in section~3.
We present our results on the orbit of XO-2 and
discussions on possible migration mechanisms of XO-2b in section~4.
Finally, we summarize the findings of this letter in section~5.

\section{Observations and Data Reductions}

We observed XO-2 with the standard I2a setup of
the High Dispersion Spectrograph (HDS: \cite{2002PASJ...54..855N})
onboard the Subaru 8.2m telescope on Mauna Kea.
An iodine gas absorption cell was inserted
for RV measurements.
We measured RVs around two full transits of XO-2b
on UT 2008 March 9 and UT 2009 November 24.
We also gathered out-of-transit RVs
spanning about two years from 2008 to 2010.
Slit width was either $0\farcs4$ ($R=90000$) or $0\farcs6$ ($R=60000$)
depending on observing conditions.
Exposure times were 600 s for around-transit phase
and 900 s for out-of-transit phase.
Finally, we obtained a high signal-to-noise ratio (SNR over 200)
and high spectral resolution ($R=90000$) template spectrum
on UT 2011 February 15 with an exposure time of 1800 s.

We process the observed frames with standard IRAF
procedures and extract 1D spectra.
Relative RVs and uncertainties
are computed by the algorithm of \citet{1996PASP..108..500B}
and \citet{2002PASJ...54..873S}.
We estimate the uncertainty of each RV
based on the scatter of RV solutions for
$\sim$4~\AA~segments of each spectrum as described in
\citet{2007PASJ...59..763N}.
The RVs and internal errors are summarized in table~1,
and plotted in the top panel of figure~1.
Note that we use HJD$_{\rm UTC}$ (Heliocentric Julian Date
in the Coordinated Universal Time) as a time standard.

Since we do not have good photometric transit data,
we additionally incorporate five published photometric
light curves taken with Keplercam on the 1.2m telescope at
the Fred Lawrence Whipple Observatory (FLWO) on Mount Hopkins, Arizona
\citep{2009AJ....137.4911F}.
Those transits were observed in the Sloan z' band.
Following the procedure by \citet{2009AJ....137.4911F} and
\citet{2010PASJ...62L..61N}, we rescale the photometric uncertainties
to account for time-correlated noise (so-called red noise:
see e.g., \cite{2006MNRAS.373..231P}) by calculating a red noise factor
$\beta = \sigma_{{\rm N,obs}}/\sigma_{{\rm N,ideal}}$
for various $N$ (corresponding to 10-20 min), and multiply the photometric
uncertainties of each dataset by the maximum value of $\beta$.
We note that we find slightly higher values of $\beta$ than
\citet{2009AJ....137.4911F}, and thus our photometric uncertainties
are conservative.

\begin{table}[htb]
\caption{RV data taken with the Subaru HDS.}
\begin{center}
\begin{tabular}{lcc}
\hline
\multicolumn{3}{c}{SUbaru HDS RVs}\\
Time [HJD\_UTC]  & Value [m~s$^{-1}$] & Error [m~s$^{-1}$]\\
\hline
2454534.71592&   0.00&   3.53\\
2454534.72354&   5.83&   3.09\\
2454534.73116&  -0.19&   3.03\\
2454534.74713&   2.47&   3.28\\
2454534.75474&  -4.44&   2.92\\
2454534.76237&  -9.72&   2.68\\
2454534.76999&  -8.87&   2.88\\
2454534.77761&  -8.62&   2.73\\
2454534.78523&  -6.95&   3.11\\
2454534.79284&  -8.66&   2.85\\
2454534.80046& -12.98&   2.82\\
2454534.80807& -19.25&   3.35\\
2454534.81570& -14.86&   2.92\\
2454534.82333& -17.81&   3.20\\
2454534.83095& -17.81&   2.76\\
2454534.83856& -21.79&   3.16\\
2454534.84618& -18.34&   3.07\\
2454534.85380& -19.09&   3.19\\
2454534.86141& -15.27&   2.89\\
2454534.86904& -16.67&   3.85\\
2454534.87665& -24.82&   3.32\\
2454534.88427& -23.42&   3.00\\
2454534.89188& -26.77&   3.12\\
2454534.89949& -37.70&   3.60\\
2454534.90711& -39.15&   3.17\\
2454534.91472& -44.97&   3.35\\
2454534.92234& -50.05&   3.42\\
2454534.92996& -51.32&   3.43\\
2454534.93759& -50.14&   3.05\\
2454534.94520& -51.64&   2.84\\
2454534.95282& -43.37&   3.53\\
2454534.96043& -48.30&   3.25\\
2454534.96805& -48.68&   3.22\\
2454616.73649& -120.61&   3.04\\
2454616.75121& -124.63&   3.02\\
2455160.01806&  -6.98&   2.87\\
2455160.02580& -10.17&   3.44\\
2455160.03354&  -2.75&   3.50\\
2455160.04127&  -0.66&   3.28\\
2455160.04900&  -3.77&   3.53\\
2455160.05675&   3.62&   3.52\\
2455160.06449&  -7.39&   3.23\\
2455160.07224&  -8.94&   2.90\\
2455160.07999& -16.07&   2.96\\
2455160.08773& -16.25&   3.23\\
2455160.09570& -28.54&   3.52\\
2455160.10344& -34.43&   2.95\\
2455160.11118& -34.92&   2.78\\
2455160.11893& -38.27&   3.51\\
2455160.12668& -35.58&   2.85\\
2455160.13442& -39.42&   2.75\\
2455160.14217& -32.60&   2.98\\
2455160.14991& -32.10&   2.83\\
2455210.99190& -45.61&   3.11\\
2455211.00660& -39.25&   2.88\\
2455211.02128& -44.23&   3.70\\
2455211.75316&  63.93&   4.56\\
2455211.76263&  71.23&   4.44\\
2455232.09437&   3.96&   3.67\\
\hline
\multicolumn{3}{l}{\hbox to 0pt{\parbox{80mm}{\footnotesize
$^{*}$ All data are presented in the electric table.
}\hss}}
\end{tabular}
\end{center}
\end{table}

\section{Model}

First we model the RVs assuming a single orbiting planet.
We then find RVs around the second transit
are vertically off by about 15~m~s$^{-1}$ from those around
the first transit (see the middle panel of figure~2).
We also find that residuals from the single planet model
has a linear trend in time (see the middle panel of figure~1).
We thus model our data assuming a single orbiting planet
with a linear RV acceleration.
The five FLWO transit light curves are
simultaneously fitted assuming the quadratic limb-darkening law.
We note that we fix one coefficient ($u_1$) of the limb-darkening
parameters to $u_1 = 0.25$, which is the central value derived by
\citet{2009AJ....137.4911F}.
This treatment is useful so as to avoid underestimate
of uncertainties for fitted parameters
(see \cite{2008MNRAS.386.1644S}).

Consequently, our model has 16 free parameters:
the RV semiamplitude $K$,
the eccentricity $e$, the argument of periastron $\varpi$,
the sky-projected stellar rotational velocity $V \sin I_{\rm{s}}$,
the sky-projected spin-orbit alignment angle $\lambda$.
the offset RV for the Subaru data $\gamma$,
the RV acceleration $\dot{\gamma}$,
the planet-star radii ratio $R_{\rm{p}}/R_{\rm{s}}$,
the orbital inclination $i$,
the semi-major axis in units of the stellar radius $a / R_{\rm{s}}$,
one of the limb-darkening coefficients $u_2$, and
the five mid-transit times of the FLWO data $T_{\rm{c}} (E)$.
We fix $P=2.6158640$ days and $T_c (0) = 2454466.88467$ in HJD$_{\rm UTC}$
which are the values reported by \citet{2009AJ....137.4911F}.
Uncertainties of mid-transit times for the two spectroscopic
transit observations (UT 2008 March 9 and UT 2009 November 24)
are well within the shortest exposure time of RVs (600 s)
and have little effect on results.

The $\chi^2$ statistic for the joint fit is
\begin{eqnarray*}
\chi^2 &=& \sum_i \left[ \frac{f_{i,{\rm obs}}-f_{i,{\rm model}}}
{\sigma_{i}} \right]^2
+ \sum_j \left[ \frac{v_{j,{\rm obs}}-v_{j,{\rm model}}}
{\sigma_{j}} \right]^2,
\end{eqnarray*}
where $f_{i,{\rm obs}}$, $v_{j, {\rm obs}}$, $\sigma_{i}$, and $\sigma_{j}$
are the observed relative fluxes, RVs, and their uncertainties.
We note that we do not add RV jitter in $\sigma_{j}$,
since reduced $\chi^2$ for the RV data is below unity
as shown later (see table~2).
The modeled fluxes ($f_{i,{\rm calc}}$) are computed by
the formula given by \citet{2009ApJ...690....1O}, and
the modeled RVs ($v_{j, {\rm model}}$) are calculated as
$v_{\rm calc} = v_{\rm Kepler} + v_{\rm RM} + \dot{\gamma}\, t + \gamma$,
where $v_{\rm Kepler}$ is the Keplerian motion,
$v_{\rm RM}$ is the RM effect, and
$\dot{\gamma}\, t$ represents the long-term RV trend.
We compute the RM effect $v_{\rm RM}$ by the accurate analytic formula
for a K0V star described in Hirano~et~al.~(2011a).
In the formula, we use $u_1=0.714$ and $u_2=0.114$ for the band of
the iodine absorption lines, based on the table of
\citet{2004A&A...428.1001C}.
We then determine optimal parameter values by minimizing the $\chi^2$
statistic using the AMOEBA algorithm \citep{1992nrca.book.....P}.
Uncertainties of free parameters are estimated by
the criterion $\Delta \chi^2 = 1.0$.

\begin{figure}[pthb]
 \begin{center}
  \FigureFile(70mm,70mm){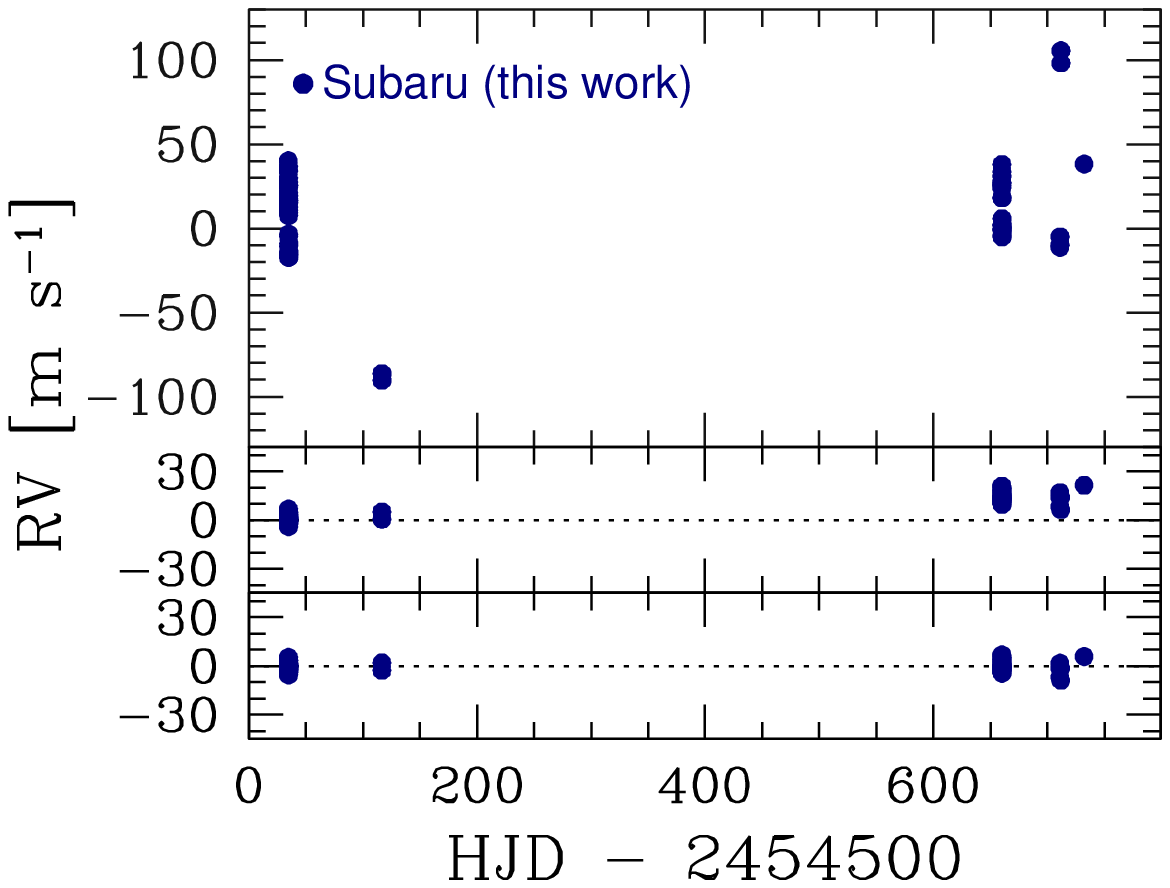}
 \end{center}
  \caption{Top panel: RVs of XO-2 observed with the Subaru HDS.
	Middel panel: Residuals of RVs from the best-fit model without
        subtracting the long-term RV trend.
	Bottom panel: Same as the middle panel but with
        subtracting the RV trend.}
 \begin{center}
  \FigureFile(70mm,70mm){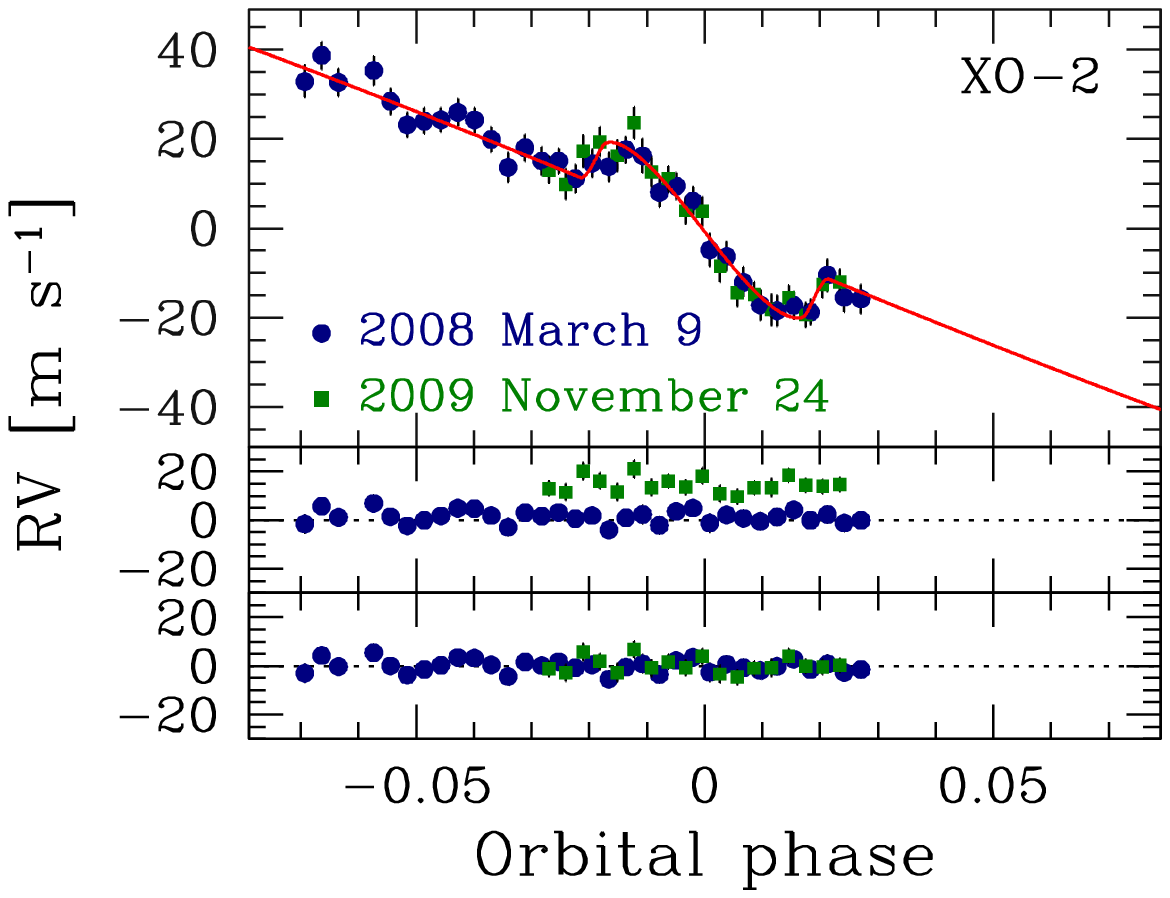}
 \end{center}
  \caption{Top panel: Phased RVs around transits after subtracting
        the RV trend. The circle and
	square symbols represent RVs on UT 2008 March 9 and
	UT 2009 November 24, respectively.
	Middel panel: Residuals of RVs from the best-fit model without
        subtracting the long-term RV trend.
	Bottom panel: Same as the middle panel but with
        subtracting the RV trend.}
 \begin{center}
  \FigureFile(70mm,70mm){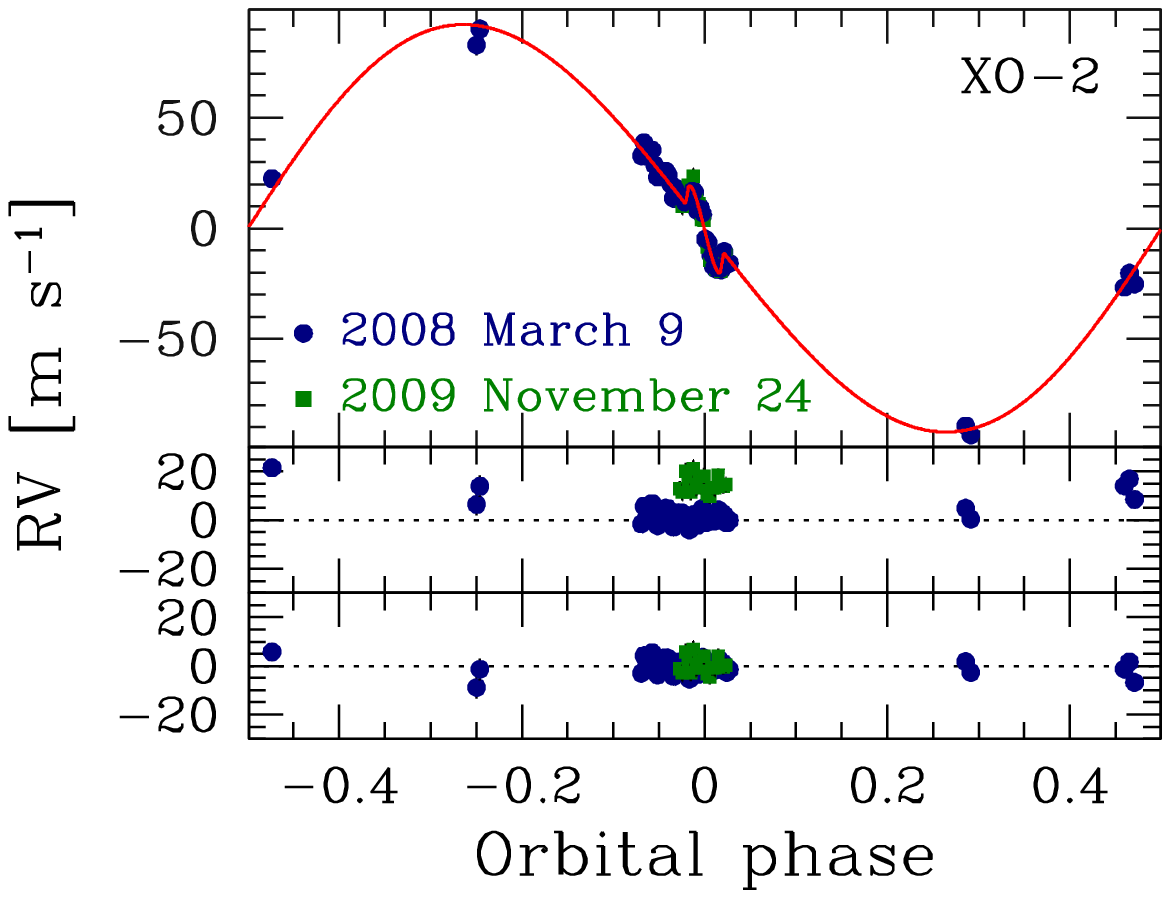}
 \end{center}
  \caption{The same as figure~2, but for a whole orbital phase.}
\end{figure}

\section{Results and Discussions}

The best-fit parameters and errors are summarized in table~2.
Photometric parameters related with only the FLWO data
($R_{\rm p}/R_{\rm s}$, $u_2$, and $T_{\rm c}$) are essentially
consistent with
the results of \citet{2009AJ....137.4911F} and are not shown.
Figure~1 plots the Subaru RVs (the top panel), residuals from
the best-fit model without (the middle panel) and with
(the bottom panel) a subtraction of the RV trend $\dot{\gamma}\, t$.
The $\chi^2$ for 59 RVs is 48.8 and the rms of RVs is 3.08~m~s$^{-1}$.
This validates our treatment of stellar jitter in the RV errors.
We find that the RV semi-amplitude $K$ ($92.2\pm1.7$~m~s$^{-1}$)
is about 8\% larger than than the value ($85\pm8$~m~s$^{-1}$)
reported by \citet{2007ApJ...671.2115B} with a smaller uncertainty.
This provides an improved estimate of the mass of XO-2b as
$0.62\pm0.02$~$M_{\rm Jup}$.
We also find the eccentricity of XO-2b is small and
consistent with zero within about 2$\sigma$.
We note that the results and errors are essentially unchanged
even if we jointly fit the RV data by \citet{2007ApJ...671.2115B}.

Figure~2 shows phased RVs around the transit phase with the best-fit model.
The RV trend is subtracted in the top and bottom panels,
but not subtracted in the middle panel.
Figure~3 is the same as figure~2, but for a whole orbital phase.
The shape of the RM effect clearly suggests a prograde orbit of XO-2.
However, the derived constraint on $\lambda$ is very poor,
namely $\lambda=10^{\circ}\pm72^{\circ}$.
It is because the orbital inclination $i$ of XO-2b is near $90^{\circ}$.
Figure~4 plots a $\Delta \chi^2$ contour map
in ($\lambda$, $V \sin I_{\rm s}$) space.
Our best-fit value of $V \sin I_{\rm s}$ based on the RM effect
is 1.45~km~s$^{-1}$, which is in good agreement with
a value $1.4 \pm 0.3$~km~s$^{-1}$
derived by \citet{2007ApJ...671.2115B} based on
a spectroscopic line analysis.
\citet{2010ApJ...719..602S} independently estimated $V \sin I_{\rm s}$
of XO-2 as $1.69 \pm 0.2$~km~s$^{-1}$ based on its mass and age,
and concluded that the XO-2 system is likely to be aligned.
The conclusion is apparently consistent with our result.
Thus although a larger value of $V \sin I_{\rm s}$ is allowed
by the RM effect, such a larger $V \sin I_{\rm s}$ is unlikely
when considering the other analyses.
As a test case, if we add a prior constraint $1.4 \pm 0.3$~km~s$^{-1}$
to the $\chi^2$ statistic, we find
$\lambda = 9^{\circ}$$^{+26^{\circ}}_{-34^{\circ}}$.
Thus the orbit of XO-2b is unlikely to be highly tilted, but
an exact value of $\lambda$ is unfortunately indeterminate.
As a result, we conclude a small eccentricity
and a likely spin-orbit alignment for the orbit of XO-2b.
As is known today, RM measurements have shown that about two thirds of
close-in giant planets are spin-orbit aligned and prograde,
whereas the last third is strongly
misaligned or even retrograde (e.g., \cite{2011A&A...527L..11H}).
Our result suggests that XO-2b is apparently in the first category.

We determine the RV acceleration of XO-2 as
$\dot{\gamma} = 7.51 \pm 0.58$~m~s$^{-1}$~yr$^{-1}$.
The trend is small but significant in the two years.
If we do not include $\dot{\gamma}$ as a free parameter,
the $\chi^2$ value for the RVs changes from 48.8 to 231.9.
Assuming a long-term linear trend,
this RV acceleration can be explained by a hypotherical third body
(index ``c'') whose mass and orbit follow a relation
$M_{\rm c} \sin i_{\rm c}/a^2_{\rm c}
= 0.042 \pm 0.003$ $M_{\rm Jup}$ ${\rm AU}^{-2}$.
The orbital period of the third body can be longer than $\sim$8yr
($\sim4$AU) since the RV trend is apparently linear spanning about 2 years.
For example, a 1~$M_{\rm Jup}$ planet at 5~AU like the Jupiter in our
Solar System meets the relation.
We note that the binary companion XO-2S does not meet this relation.
There is another possibility that the timescale of the additional RV variation
is a few hundred days, due to the lack of our observations between
the two transits. Additional RV measurements with the Subaru HDS will
allow us to discriminate the timescale of the RV variation.

Other possibilities of the trend include a systematic RV variation of the
Subaru HDS, starspots 
or magnetic cycles of XO-2.
The first scenario is unlikely,
because RVs of the Subaru HDS for RV standard stars are stable
within a few m~s$^{-1}$ \citep{2010ApJ...715..550H,2009ApJ...703..671S}.
The starspot scenario is also unlikely, because
no significant RV variation has been observed for XO-2
in a timescale corresponding to the rotation period of
the host star ($\sim$35 day: \cite{2010ApJ...725.1995M}),
although RV variations caused by spots should have been maximum.
At this point, we cannot exclude the possibility of magnetic cycles
of XO-2 recently examined by \citet{2011arXiv1107.5325L},
although any strong magnetic activity has not been reported for this system.
To specify the cause of the RV variation and make a decisive conclusion,
further RV monitoring would be very important in this system.

There is still a possibility that the mutual inclination between
the orbital axes of the transiting planet XO-2b and the binary
companion XO-2S was once larger than the threshold value of
the Kozai mechanism.
In addition, if the eccentricity of the binary orbit is larger
than $\sim$0.9, the timescale of the Kozai migration might have been
shorter than that of General Relativity for XO-2b at the birthplace
($\sim$2Gyr based on \cite{2003ApJ...589..605W}).
However, the presence of the hypotherical third body completely
dictates that XO-2b cannot be migrated through the Kozai migration
caused by XO-2S,
since the gravitational perturbation timescale due to the third body
(comparable to the orbital period of the third body) is
much shorter than the Kozai migration timescale
(see e.g., \cite{1997AJ....113.1915I, 2010PASJ...62..779N}).

On the other hand, \citet{2010ApJ...718L.145W} pointed out that
a hot Jupiter around a cool (less than $\sim$6250K) star can lead
to a re-alignment of its spin-orbit alignement angle due to the
tidal force on a convective surface layer of the cool host star.
XO-2 is in the category of a cool star
($\sim$5340K: \cite{2007ApJ...671.2115B})
and the spin axis of the host star might have been re-aligned.
Thus one possible migration scenario is that
XO-2b has migrated through planet-planet scattring
(e.g., \cite{1996Sci...274..954R, 2008ApJ...678..498N, 2008ApJ...686..580C}),
and subsequently
the eccentricity and the spin-orbit alignment angle have been damped.
In that case, the hypothetical third body may be the counterpart of
the planet-planet scattering.
Another possible migration mechanism is 
the standard disk-planet interaction mechanism
(e.g., \cite{1996Natur.380..606L, 2004ApJ...616..567I}).
Although we cannot specify the migration mechanism of XO-2b
at this point in time, further measurements of transit timings of XO-2b
would be useful to constrain another low-mass body in
mean motion resonance, which can be an evidence of
the disk-planet migration
(\cite{2009AJ....137.4911F}).
Any detection or constraint of such a mean motion resonance body
by transit timings, as well as further RV measurements, would allow
further discussions on the migration history of this system.

\begin{table}[pthb]
\caption{Best-fit values and errors of the parameters.}
\begin{center}
\begin{tabular}{l|cc}
\hline
\hline
Free Parameter & Value & Error\\
\hline
$K$ [m s$^{-1}$] 
& 92.2 & $\pm 1.7$ \\
$e$ 
& 0.045 & $\pm 0.024$ \\
$\varpi$ [$^{\circ}$]
& 270 & $^{+10}_{-7}$ \\
$V \sin I_{\rm s}$ [km s$^{-1}$]
& 1.45 & $^{+2.73}_{-0.14}$ \\
$\lambda$ [$^{\circ}$]
& 10  & $\pm 72$\\
$i$ [$^{\circ}$]
& 88.7 & $^{+1.3}_{-0.9}$ \\
$a / R_{\rm s}$
& 8.43 & $^{+0.32}_{-0.34}$ \\
$\gamma$ [m s$^{-1}$] 
& -34.3  & $\pm 1.1$\\
$\dot{\gamma}$ [m s$^{-1}$ yr$^{-1}$] 
& 7.51 & $\pm 0.58$ \\
\hline
\hline
Derived Parameter & Value & Error\\
\hline
$M_{\rm p}$ [$M_{\rm Jup}$] 
& 0.62  & $\pm 0.02$ \\
$M_{\rm c} \sin i_{\rm c}/a^2_{\rm c}$ [$M_{\rm Jup}$ ${\rm AU}^{-2}$]
& 0.042 & $\pm 0.003$ \\
$\chi^2$ for 59 RVs
& 48.8  & -- \\
rms [m s$^{-1}$] 
& 3.08  & -- \\
\hline
\multicolumn{3}{l}{\hbox to 0pt{\parbox{100mm}{
}\hss}}
\end{tabular}
\end{center}
\end{table}

\begin{figure}[pthb]
 \begin{center}
  \FigureFile(65mm,65mm){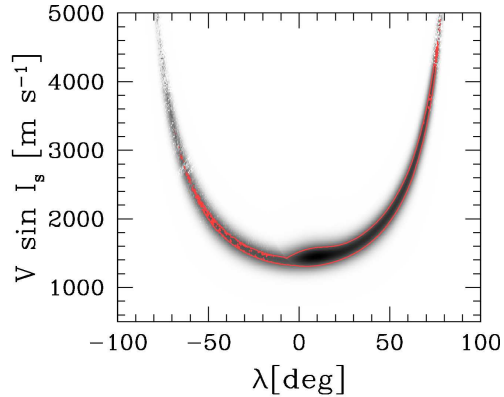}
 \end{center}
  \caption{
  A $\chi^2$ contour map in $\lambda$-$V \sin I_s$ space.
  The solid lines represent contour for $\Delta \chi^2 = 1.0$.}
\end{figure}

\section{Summary}

We have monitored RVs of XO-2 spanning about two years with the Subaru HDS.
We find a small eccentricity and a likely spin-orbit alignment for
the orbit of the transiting planet XO-2b, and we also detect
a long-term RV acceleration of the host star.
Based on the observed properties of this system,
we constrain the Kozai migration scenario of XO-2b.
This illustrates that a presence of a wide binary companion does not
necessarily suggest the Kozai migration (see also
\cite{2009PASJ...61L..35N,2009ApJ...703L..99W,2010PASJ...62..779N},
for the migration mechanism of a retrograde planet HAT-P-7b).
To constrain the occurrence frequency of the Kozai migration
for planets in binary systems, combinations of observations of
(1) the orbital eccentricity by RV measurements,
(2) the spin-orbit alignment angle by the RM effect,
and (3) the binarity (or multiplicity) of the system by
high-contrast direct imaging, are of special importance.
Accumulating those observations for transiting planetary systems
would allow us to learn more about planetary migration
mechanisms in the future.

\bigskip
This letter is based on data collected at Subaru Telescope,
which is operated by the National Astronomical Observatory of Japan.
We acknowledge a kind support by Akito Tajitsu
for the Subaru HDS observations.
We are grateful for Zach Gazak for helpful discussions on
XO-2 photometric observations.
The data analysis was in part carried out on common use data analysis
computer system at the Astronomy Data Center, ADC,
of the National Astronomical Observatory of Japan.
N.N. acknowledges a support by NINS Program for Cross-Disciplinary Study.
T.H. is supported by a Japan Society for Promotion of Science
(JSPS) Fellowship for Research (DC1: 22-5935).
M.T. is supported by the Ministry of Education, Science,
Sports and Culture, Grant-in-Aid for
Specially Promoted Research, 22000005.
We wish to acknowledge the very significant cultural role
and reverence that the summit of Mauna Kea has always had within
the indigenous people in Hawai'i.



\end{document}